\documentclass[aps,prb,preprint,groupedaddress,showpacs]{revtex4}

\usepackage{graphicx}
\usepackage{epsfig}

\def\gamnas{{Ga$_{1-x}$Mn$_{x}$As}}

\begin{document}


\title{Optimizing $T_c$ in the (Mn,Cr,Ga)As and (Mn,Ga)(As,P) Ternary Alloys}



\author{J.~L. Xu}
\affiliation{Department of Chemical and Materials Engineering,
 Arizona State University, Tempe, AZ, 85287}
\email[]{Mark.vanSchilfgaarde@asu.edu}

\author{M. van Schilfgaarde}
\affiliation{Department of Chemical and Materials Engineering,
 Arizona State University, Tempe, AZ, 85287}

\date{\today}

\begin{abstract}

We explore two possible ways to enhance the critical temperature
$T_c$ in the dilute magnetic semiconductor
Mn$_{0.08}$Ga$_{0.92}$As.  Within the context of the
double-exchange and RKKY pictures, the ternary alloys
Mn$_{x}$Cr$_{0.08-x}$Ga$_{0.92}$As and
Mn$_{0.08}$Ga$_{0.92}$As$_y$P$_{1-y}$ might be expected to have
$T_c$ higher than the pseudobinary Mn$_{0.08}$Ga$_{0.92}$As.  To
test whether the expectations from model pictures are confirmed,
we employ linear response theory within the local-density
approximation to search for theoretically higher critical
temperatures in these ternary alloys.  Our results show that
neither co-doping Mn with Cr, nor alloying As with P improves
$T_c$.  Alloying with Cr is found to be deleterious to the $T_c$.
Mn$_{0.08}$Ga$_{0.92}$As$_y$P$_{1-y}$ shows almost linear
dependence of $T_c$ on $y$.

\end{abstract}


\maketitle

Searching for spintronic materials with high Curie temperature
$T_c$ has attracted a lot of interest recently \cite{Ohno96,
Ohno99} because of the promising future of these materials.  In
spintronics technology, by including both magnetic and electronic
properties in the devices, we can manipulate the electronic spin
to the same degree as what we do to the electronic charge in
conventional semiconductor devices.  It can greatly improve the
performance of current digital information processing and storage
technology.  Currently, one of the central issues for spintronics
application is the development of carrier-mediated ferromagnetic
semiconductors that are magnetic above room temperature.  Dilute
magnetic semiconductors (DMS), i.e. semiconductors doped with low
concentrations of magnetic impurities (such as Mn, Co, Cr), are
generally thought to be good candidates for these spintronics
applications. \gamnas\ is one of the most widely studied DMS,
because it is one of the few materials where it is generally
agreed that the magnetism is carrier-mediated.

Currently, much effort has been expended to increase $T_c$ in
\gamnas\ and related DMS materials.  $T_c$ has risen steadily in
\gamnas, to $T_c\sim{}180$K in very thin films, carefully
annealed to eliminate Mn interstitials without at the same time
allowing MnAs precipitates to form\cite{Chiba03, Ku03, Edmonds04}.
However, there is still a large gap between the highest $T_c$
obtained to date and a technologically useful working
temperature.

A systematic understanding of the ferromagnetism in DMS systems
is necessary for the design and optimization of the materials.
However, the theoretical understanding of this system is still
not very clear, although several models were proposed.  One
generally used model is based on the assumption that Mn local
magnetic moments interact with each other via RKKY-type
interactions, the Mn at the same time providing a source of free
holes\cite{Dietl00}.  This picture can explain the temperature
dependence of the magnetization and the hole resistivity of the
DMS, although there is a growing consensus that many of the
claims of that paper were artifacts of the assumptions in the
model.  On the other hand, Akai\cite{Akai98} first used the local
spin-density approximation (LSDA) to estimate $T_c$ within the
Coherent Potential Approximation (CPA) in (In,Mn)As; he argued
that a double exchange mechanism was a more appropriate
description of the magnetism than the $pd$ exchange assumed by
Dietl.

In addition, there is a fair amount of research that indicates the
ferromagnetism in DMS systems are strongly affected by the
inherent disorder.  In our recent work\cite{Xu05} we included
various effects of disorder, and found that there is a rather
strict upper limit of around 250K for $T_c$ in disordered Mn:GaAs
systems (as computed within the LDA), which implies a rather
unpromising future for the simple Mn:GaAs alloy.  So, it is
becoming an important issue to find possible ways to optimize the
magnetic properties of DMS.  The possible means may include
co-doping the magnetic elements, change of host materials and
magnetic elements, or spacial confinement of the magnetic dopants,
etc.  We use the Zener double-exchange picture for guidance, for
reasons that will be clear below.  For detailed analysis we employ
to the same LDA linear response techniques as in Ref.~\cite{Xu05}.


\begin{figure}[ht]
\centering
\includegraphics[width=6cm]{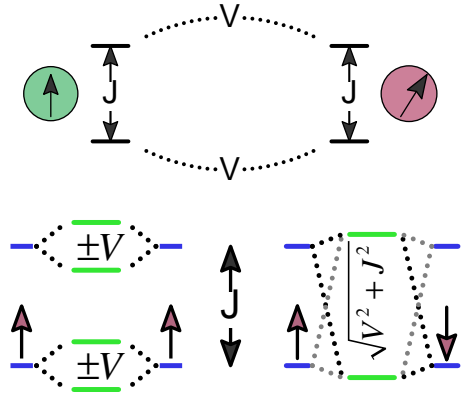}
\caption{Cartoon of the Zener model coupling levels on two
  neighboring sites.  Top illustrates the majority and minority
  levels on each site together with the model parameters
  (intra-atomic exchange splitting $J$, effective interatomic
  hopping matrix element $V$, and canting angle $\theta$).
  Bottom left depicts the four levels when spins are aligned parallel;
  bottom right depicts the levels when spins are aligned antiparallel.}
\label{fig:zenermodel}
\end{figure}

Here we consider two strategies to make alloys related to
Mn$_{x}$Ga$_{1-x}$As, that can have higher $T_{c}$.  We based both
strategies on the double exchange/superexchange
picture\cite{Anderson55} depicted in Fig.~\ref{fig:zenermodel}.
The four levels are described by the hamiltonian
\begin{equation}
H \ = \ U^{\dagger} \
\left(
\begin{array}{cccc}
 +J   &  0   & \ V   &  0 \\
  0   & -J   & \ 0   &  V \\
  V   & 0    & +J    &  0 \\
  0   & V    &   0   & -J
\end{array}
\right) U,
\end{equation}
with $U$ the spinor rotation matrix for the second site:
\begin{equation}
U \ = \
\left(\begin{array}{rr}
       1  & 0   \\
       0  & U_2 \\
      \end{array}
\right) ,
\quad
U_2\ =\
\left(\begin{array}{rr}
      \cos\theta/2 & {\rm{-}}\sin\theta/2 \\
      \sin\theta/2 &  \cos\theta/2 \\
      \end{array}
\right)
\end{equation}
The eigenvalues are given by the quartic equation
\cite{Anderson55}
\begin{equation}
\epsilon^2 = J^2 + V^2 \pm 2 V J \cos(\theta/2) .
\end{equation}
When only the lowest state is filled, the parallel alignment
dominates (Zener double exchange).  The most salient other
characteristic of the Zener double exchange energy is its unusual
non-Heisenberg dependence on angle, varying as
$|\cos(\theta/2)|$.  When the first two states are filled the
energy vanishes for parallel spin alignment, and the the system
is stabilized in the AFM alignment (Anderson superexchange), the
energy varying as $-\cos(\theta)$.  As a function of filling,
there is a transition from FM stabilization to AFM.  The maximum
FM stabilization occurs at half-filling.

In the Mn$_{x}$Ga$_{1-x}$As and Cr$_{x}$Ga$_{1-x}$As alloys, the
exchange is mediated by the partially filled TM-derived $t_2$
impurity band.  In the dilute limit, the Mn-derived $t_2$ level
has threefold degeneracy and falls at VBM+0.1~eV, while the
Cr-derived level falls at VBM+0.3~eV.  These levels broaden into
an impurity band as ${x}$ increases.  In reality this band is
composed of an admixture of TM $d$ states and effective-mass like
states of the host\cite{mark01,Mahadevan04}, mostly derived from
the top of the host valence band.

To the extent that the model correctly describes exchange in
Mn$_{x}$Ga$_{1-x}$As, it is possible to optimize $T_c$ of a
specific DMS system by varying the carrier concentration to set
this impurity band as close to half filling as possible.  One
possible way to do this is to make a pseudoternary alloy
Mn$_{x}$Cr$_{y}$Ga$_{1-x-y}$As, where the (Mn,Cr,Ga) atoms share
a common (fcc) sublattice.  Cr has one electron to fill the
3-fold majority $t_2$ level to 1/3 filling.  Mn has two
electrons, so the majority $t_2$ level should be 2/3 full.  By
alloying Mn and Cr, we might obtain a compromise 1/2 filling that
will increase $T_{c}$.

Another possibility is to vary the anion in the III-V series. From
model viewpoints (and confirmed by the LDA), there are two
competing effects from the anion.  The important quantity is the
position of the transition-metal $d$ level relative to the top of
the valence band (determined largely by the anion $p$ level). When
the anion is heavy (e.g. Sb) the $p$ is shallow; the Mn $d$ sits
mostly below the valence band, and the $t_2$ derived impurity band
is mostly composed of the effective-mass like host states; thus
the exchange is effectively mediated the holes near the valence
band top, as the RKKY picture assumes. If the Mn $d$ is very deep,
the coupling between the Mn and holes also weak, so the
interatomic exchange is small.  As the anion becomes lighter
(Sb$\to$As$\to$P$\to$N) the anion $p$ deepens, so that finally the
Mn $d$ falls above the valence band and forms a deep level.  In
this limit the coupling should be well described by
double-exchange. For Mn in GaN, the $t_2$ level is approximately
at midgap. It couples strongly to nearest neighbors, but is short
ranged.  At least in the dilute case, the near-neighbor exchange
interactions are mostly ferromagnetic and rather strong, but they
decay rapidly with distance \cite{Xu05}. But also in the dilute
case, there are few nearest neighbors and more distant
interactions are necessary for long-range order to exist, as
without them there is no percolation path.  The net result is that
$T_c$ is predicted to be quite low in Mn$_{x}$Ga$_{1-x}$N
\cite{Sato04}.  The anion with the highest $T_c$ will adopt some
compromise between a very deep valence band (GaN) and a very
shallow one (GaSb).  With this picture, and the expectation that
the Mn $d$ level in MnGaP is perhaps already too high, we might be
able to make a ternary MnGaP$_{x}$As$_{1-x}$ where ${x}$ can be
adjusted to tune the position of the valence band, and thus
optimize $T_{c}$.

First principles calculations provide a way for evaluation of the
exchange interactions and $T_{c}$ in a DMS system of a specific
geometry.  We adopt here the same scheme as presented in
Ref.\cite{Xu05}, to study the ferromagnetic trends for the
ternaries Mn$_{x}$Cr$_{0.08-x}$Ga$_{0.92}$As and
Mn$_{0.08}$GaP$_{y}$As$_{1-y}$.  Self-consistent calculations of
200-atom special quasirandom structures\cite{zunger90}, or SQS,
were performed using the method of Linear-Muffin-Tin Orbitals
(LMTO) within the local spin density approximation (LSDA) and the
Atomic Spheres Approximation\cite{licht87,mark99}.  The exchange
interactions were computed within the long-wave approximation
using a linear response Green's function technique, which maps the
total energy onto a Heisenberg form. The interatomic Heisenberg
exchange parameters are determined from the linear response
expression
\begin{equation}
J_{ij}^{\perp} =\frac{1}{2\pi}\int^{\varepsilon _{F}}d\varepsilon
{\rm \,ImTr}_{L}\left\{ P_{i}\cdot \left( T_{ij}^{\uparrow}\cdot
T_{ji}^{\downarrow}+T_{ij}^{\downarrow}\cdot T_{ji}^{\uparrow}\right)
\cdot P_{j}\right\}
\end{equation}
where $i$ and $j$ are indexes of atoms with the on-site
perturbation having a form
$P=P_{Z}=(T_{\uparrow}^{-1}-T_{\downarrow }^{-1})/2=(P^{\uparrow
}-P^{\downarrow })/2$, as described in Ref.\cite{mark99}.  $P$ is
a diagonal matrix whose rows and columns correspond to the
orbitals at a particular site ($spd$ in the present case); $P$ is
closely related to the majority-minority spin-splitting.
$T_{ij}^{\uparrow}\cdot T_{ji}^{\downarrow}$ essentially the
transverse susceptibility.  Magnetization at finite temperature
and $T_{c}$ were calculated using the Cluster Variation Method
(CVM) of Kikuchi\cite{kikuchi}, generalized to the continuous
rotational degrees of freedom of the (classical) Heisenberg model,
which predicts $T_{c}$ to within about 5\% of results calculated
by full spin-dynamics simulations in these DMS alloys\cite{Xu05}.
For both Ga$_{0.92}$Mn$_{0.08}$As and Ga$_{0.92}$Cr$_{0.08}$As the
CVM predicts $T_c$ of about 280K. We will show that, although
these two systems are intriguing based on the simple pictures we
carry around about them, neither alloy is predicted to have
magnetic properties more favorable than Mn$_{x}$Ga$_{1-x}$As at
$x=0.08$.

Before addressing $T_c$ in the pseudoternary
Mn$_{x}$Cr$_{y}$Ga$_{1-x-y}$As, we consider a simple model
calculation to see to what extent the LDA exchange interactions
approximately parallel the double-exchange picture.  Starting
from Mn$_{0.08}$Ga$_{0.92}$As SQS configuration, we artificially
shift the Fermi level of this system to mimic additional
doping. In this model dopant electrons are assumed to have no
effect but change the filling of the valence band, which enables
us to isolate this effect from other complexities that occur in
the real system, but which are missing in the double-exchange
model.

\begin{figure}[ht]
\centering
\includegraphics[width=8cm]{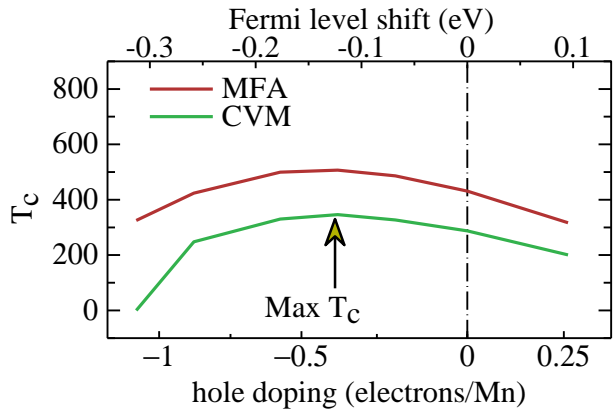}
\caption{Dependence of $T_c$ Fermi level shift and the
corresponding change in band occupancy. Calculated MFA and CVM
$T_c$ are presented as a function of Fermi level shift and
deviation from charge neutrality. }
\label{fig:fermi}
\end{figure}

Fig.~\ref{fig:fermi} shows the change in $T_c$ due to the Fermi level
shift $\Delta E_F$ (and corresponding hole doping), in a 200-atom SQS structure
Ga$_{92}$Mn$_{8}$As$_{100}$.  As predicted by both the RKKY
picture and the double-exchange picture, increasing $E_F$ reduces
the number of holes and reduces $T_c$. Conversely, with the
initial introduction of more holes ($\Delta E_F < 0$) $T_c$
increases as both RKKY picture and double-exchange models predict.
However, fig.~\ref{fig:fermi} shows that $T_c$ reaches a maximum
for $\Delta E_F\sim-0.20$~eV; further shifts reduce $T_{c}$.
Observing the Fermi level shifts in the DOS
we can confirm that position of $E_{F}$ at optimum $T_{c}$
corresponds to $E_{F}$ falling in the middle of the Mn $t_2$ band,
indicating half filling of the $t_2$ level. This can be determined
independently by counting the number of holes added as a function
of $\Delta E_{F}$.  $\Delta E_F\sim-0.20$~eV corresponds to the
addition of approximately 4 holes in this system of 8 Mn atoms, or
0.5 hole/atom.  The $t_2$ level with initially 2 electrons
($\Delta E_{F}=0$) changes to 1.5 electrons near $\Delta
E_{F}=-0.2$, which is just half of the number needed to fill the
level.  This confirms that the double-exchange picture more
closely reflects the {\em ab initio} LDA calculations than does
RKKY, as Akai first proposed\cite{Akai98}.

One possible way to introduce holes in a real system is to add
another dopant, e.g. Be, as has been done in $\delta$-doped
MnGaAs \cite{Wojtowicz03} However, Be doping appears to catalyze
the formation of other deleterious defects (probably Mn
interstitials, which being donors, bind to Be).  Another way is
to alloy Cr with Mn.  In the simplest view we can mimic this
effect by a ``virtual crystal'' where we use a hypothetical atom
X with atomic number 24.5.  The $T_c$ evaluated for this
hypothetical Ga$_{92}$X$_{8}$As$_{100}$ SQS structure is
dramatically larger than $T_{c}$ for Ga$_{92}$Mn$_{8}$As$_{100}$
or Ga$_{92}$Cr$_{8}$As$_{100}$, as we found before.  For this 8\%
doping case, the $T_{c}$ is calculated to be 710K with the MFA
and 390K within the more reliable CVM.

\begin{figure}[ht]
\centering
\includegraphics[width=8cm]{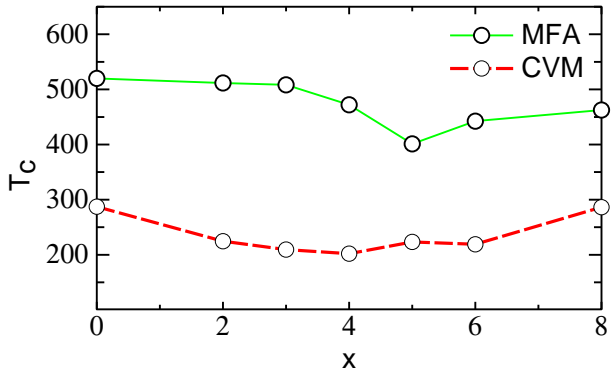}
\caption{Dependence of $T_c$ on $y$ in
Mn$_{x}$Cr$_{y}$Ga$_{1-x-y}$As. $T_{c}$ is computed both within
the MFA and the CVM.  The noise in the data occurs because $T_c$
depends on the specific choice of quasi random configuration.  For
the same $y$, there can be a deviation between different SQS
structures as much as 50K. } \label{fig:codope}
\end{figure}

Finally, we consider a physically realizable configuration,
Mn$_{x}$Cr$_{y}$Ga$_{1-x-y}$As.  We now construct a series of
three-body 200-atom SQS structures, varying $x$ for fixed
$x+y$=8\%.  $T_c$ was calculated for all a variety of
configurations with different $x$.  The calculated Curie
temperature depending on $x$ are shown in Fig.~\ref{fig:codope}.
The picture that emerges is, unfortunately very different from
the virtual crystal.  Co-doping actually decreases $T_c$.  We can
see although CVM and MFA $T_c$ curves have a little different
shape, they both predict $T_c$ to decrease near $x=1/2$, in
contradistinction to the virtual crystal.  The reason can be
traced to the fact that Cr and Mn have their impurity levels at
different energies. Consider the limit of infinitely narrow
impurity bands: there is a Cr-derived level at VBM+0.3~eV and a
Mn-derived level at VBM+0.1~eV. The single electron in the
Cr-derived level falls to the Mn level, thus emptying out the Cr
$t_2$ state and completely filling the Mn $t_2$ state.  In the
double-exchange picture, the system will be antiferromagnetic.
In the alloy, these levels broaden into bands.  However the Mn
and Cr like states derive from sufficiently different energies,
so that rather than forming a single wide impurity band made of
equal admixtures of Cr and Mn, the band is composed of a lower
part more of Mn character and an upper part dominated by Cr.  If
the bandwidth is not sufficiently large to counter this effect,
the net result is that $T_{c}(x)$ in the true ternary apparently
bows oppositely to what occurs in the idealized case.

\begin{figure}[ht]
\centering
\includegraphics[width=8cm]{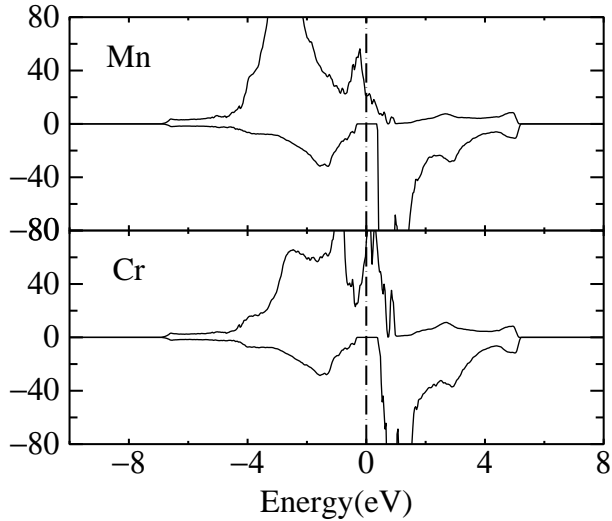}
\caption{Energy resolved Partial DOS of Mn and Cr ions in
Ga$_{92}$Mn$_4$Cr$_4$As$_{100}$, showing similar filling as single
element doping case. } \label{fig:pdos}
\end{figure}

This is reflected in the partial DOS of TM in (Mn,Cr):GaAs(Fig.
\ref{fig:pdos}).  The DOS of the $t_2$ for each individual TM ion
reflects the DOS in the respective pseudobinary alloy: the Mn
$t_2$ near $E_F$ is centered slightly below the Cr $t_2$, showing
the localized behavior of magnetic dopants.  Some broadening is
evident (particularly in the tail of the Mn $t_2$ above $E_F$,
but owing to the dilute alloy concentration the coupling between
$t_2$ levels is not sufficient to broaden it into a homogeneous
band.  The lack of this distinction makes the virtual crystal
approximation an unsuitable description of the real alloy.  When
both levels are present the deeper Mn $t_2$ (2/3 full in Mn:GaAs)
becomes fuller still, while the Cr $t_2$ (1/3 full in Cr:GaAs)
becomes more empty.  Hence, the co-doping actually drive the
filling opposite to the desired direction.  The band is actually
adversely affected, resulting in a decrease in $T_c$ by
co-doping.

It is notable that an RKKY-like picture\cite{Dietl00}, which
describes exchange interactions purely in terms of spin-split
host bands whose splitting is determined by the local magnetic
moments, will also miss this effect.  This establishes in another
way the fact that exchange interactions within the LDA more
closely reflect the Zener double-exchange model than an RKKY-like
model.


Our other strategy to increase $T_{c}$ is to consider the ternary
Mn$_{0.08}$Ga$_{0.92}$As$_y$P$_{1-y}$, as we described above.  In
this calculation, SQS configurations are also generated through
independent sublattices of cations (Mn,Ga) and anions (As,P). the
Mn concentration is fixed to be 0.08\% as before.  $T_c$ for
different P concentrations are studied.  For the lattice constant,
we use virtual crystal approximation with modification based on
Ref. \cite{Shih85}. The figure shows that the net effect
(deviation from linear behavior) is small. but in any case $T_c$
bows downward with $y$, with $T_c$ suppressed by $\sim$50K for
$y=0.5$. This reduction can be understood as a suppression,
through disorder, of the number of paths that connect different
magnetic sites, and thus the magnetic exchange between them.

\begin{figure}[ht]
\centering
\includegraphics[angle=0,width=.45\textwidth,clip]{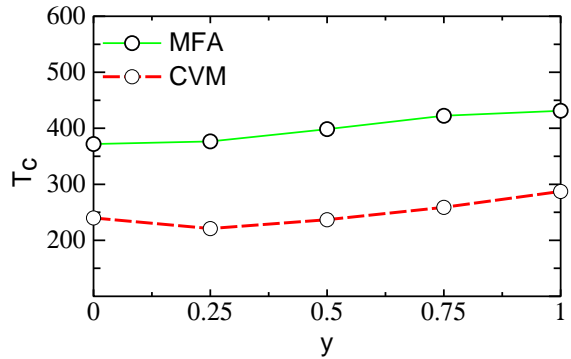}
\caption{Curie Temperature for
Mn$_{0.08}$Ga$_{0.92}$As$_y$P$_{1-y}$ as a function of $y$}
\label{fig:asp}
\end{figure}

In conclusion, based on the LSDA, we found out the Curie
Temperature of DMS can be improved by modifying the band filling,
as both double exchange and RKKY models predict.  We show that
the LSDA results more closely parallel the double exchange than
the RKKY picture, and predict that optimal $T_c$ occurs when the
impurity $t_2$ band is half full.  However, when we attempt to
modify the concentration in a real crystal like Mn and Cr
co-doping GaAs, $T_c$ is actually suppressed.  For a successful
application of this rule, it will be necessary to find another
way to to tailor the band occupancy in order to achieve half
filling.  We also found that admixing of As and P on the anion
sublattice does not improve $T_c$.


This work was supported by the Office of Naval Research.
\bibliography{dms}

\end{document}